\documentclass{ws-ijmpd}

\usepackage{lscape}
\newcommand{\be}{\begin{equation}}
\newcommand{\ee}{\end{equation}}
\newcommand{\bea}{\begin{eqnarray}}
\newcommand{\eea}{\end{eqnarray}}

\usepackage[super,compress]{cite}
\begin{document}

\markboth{L\'opez-Corredoira \& Calvo-Torel}
{Fitting of supernovae without dark energy}

%
\catchline{}{}{}{}{}
%

\title{Fitting of supernovae without dark energy}

\author{M. L\'opez-Corredoira$^{1,2}$ and J. I. Calvo-Torel$^2$}

\address{$^1$ Instituto de Astrof\'\i sica de Canarias,\\ 
E-38205 La Laguna, Tenerife, Spain\\
E-mail: martin@lopez-corredoira.com \\
$^2$ Dep. de Astrof\'\i sica, Universidad de La Laguna,\\
E-38206 La Laguna, Tenerife, Spain}

\maketitle

\begin{history}
\received{Day Month Year}
\revised{Day Month Year}
\comby{Managing Editor}
\end{history}

\begin{abstract}
With data from Pantheon, we have at our disposal a sample
 of more than a thousand supernovae Ia covering a wide range of redshifts 
with good precision. Here we make fits to the corresponding 
Hubble--Lema\^itre diagram with various cosmological models, 
with intergalactic extinction, evolution of the luminosity of 
supernovae, and redshift components due to partially non-cosmological factors. The data
are well fitted by the standard model to include dark energy, 
but there is a degeneracy of solutions
with several other variables. 
Therefore, the Hubble--Lema\^itre diagram of SNe Ia
cannot be used alone to infer the existence of the 
accelerated expansion scenario with dark energy.

Within this degeneracy,  models that give good 
fits to the data include the following alternative solutions:
Einstein--de Sitter with gray extinction $a_V=1.2\times 10^{-4}$ Mpc$^{-1}$;
linear Hubble--Lema\^itre law static Euclidean with 
gray extinction $a_V=0.4\times 10^{-4}$ Mpc$^{-1}$;
Static Euclidean with tired light and gray extinction 
$a_V=2.8\times 10^{-4}$ Mpc$^{-1}$;
Einstein--de Sitter with absolute magnitude evolution
 $\alpha =-0.10$ mag Gyr$^{-1}$;
Friedmann model with 
$\Omega _M=0.07 - 0.29$, $\Omega _\Lambda =0$ and 
partially non-cosmological tired-light redshifts/blueshift 
with attenuation/enhancement
$|K_i|<2.2\times 10^{-4}$ Mpc$^{-1}$
(although requiring calibration of $M$ incompatible with 
local SNe measurements).
\end{abstract}



\section{Introduction}
\label{S:1}
\unskip

In 1998 and 1999 Riess et al.\cite{Riess1998} and 
Perlmutter et al.\ \cite{Per98} published papers
establishing the existence of an accelerated expansion 
from the analyses of type Ia supernovae (SNe Ia).
Their sample contained few stars; for instance, Riess et al.\ 
analyzed data from 10 type Ia supernovae (SNe Ia) 
with redshifts between 0.16 and 0.62.
Along with earlier work, they augmented the sample 
with 34 other nearby supernovae and 16 more distant ones.
The luminosity distance measured with supernovae 
exceeded by up to 0.25 to 0.28 magnitudes that expected in 
the standard model in the early 1990s (open 
universe with matter density $\Omega_M \sim 0.2$). 
This excess could be explained by adding a positive 
cosmological constant.
They saw that the fit determined an accelerating 
expanding universe [$q_0 \leq 0$;
$q_0 \equiv \frac{\Omega_M}{2} - \Omega_\Lambda$] 
with a  confidence level in the range 99.5--99.9 percent.
The best fits gave the value of the parameters 
$\Omega_M = 0.24_{-0.24}^{+0.56}$, 
$\Omega_\Lambda = 0.72_{-0.72}^{+0.48}$, and  
$H_0 = 65.2 \pm  1.3$ km s$^{-1}$ Mpc$^{-1}$, 
where the newly introduced $\Omega_\Lambda \ne 0$ 
stands for a cosmological constant or quintessence component.
With more recent, larger, and more accurate sample of 
data by Ref.\ \citen{Scolnic2018} (Pantheon data), the results of Riess et al.\ 
and Perlmutter et al.\ for a universe with 
a positive cosmological constant are confirmed with much lower error bars (see 
\S \ref{.data}).

There are other ways to obtain a fit to the data 
of SNe Ia. We now analyze the different fitting 
alternatives, focusing on scenarios where no $\Lambda $ term is included.

\section{SNe Ia data: Pantheon}
\label{.data} 

The 'Pantheon' sample of SNe Ia \cite{Scolnic2018} 
is currently one of the largest samples with  highest redshift 
supernova data. Its redshift range spans between 
$z=0.001$ and $z=2.3$,  thus extending the range far beyond $z=1$. We
use the 2018  version of this sample with 1048 SNe. 
Just after we did our analyses, a more recent version was
released (Pantheon+) \cite{Sco21,Bro22}, 
with 1550 SNe Ia, most of the new sources being
 low $z$ SNe; the maximum redshift is not higher. 
In any case, for our purposes, the Pantheon sample 
is more than enough to test those cosmological 
scenarios that are more sensitive to the high $z$ sources.

Unlike previous samples where the $\Delta M$ errors 
(where $M$ is the assumed absolute magnitude of SNe Ia) 
were given, Pantheon provides only the redshifts and 
the distance moduli with their uncertainty along with 
the covariance matrix. Here we simply use the redshifts 
and apparent magnitudes of SNe Ia to produce
the fits detailed in the following sections.
There is some discussion on the discrepancies of 
redshifts and magnitudes of Pantheon with other
catalogues and on how systematic errors in redshift or 
magnitude measurements can affect the
measured parameters \cite{Ste20,Ram21}. We do 
not enter in that discussion here; we simply 
assume that the numbers in the Pantheon table 
are correct, and we question whether other 
cosmological scenarios without dark energy 
can reproduce these data with the same accuracy as
the official published results of Pantheon SNe analyses.

\section{Fit with the standard $\Lambda $CDM model}

By knowing the intrinsic luminosity $\mathcal{L}$ of a 
supernova and measuring the observed flux $\mathcal{F}$, 
we can obtain an estimate of the distance-luminosity to each of these supernovae.
\begin{equation} \label{eq1}
D_L = \left( \frac{\mathcal{L}}{4 \pi \mathcal{F}} \right)^{(1/2)}
\end{equation}

Considering as cosmological parameters the Hubble 
constant, $H_0$, the mass density, $\Omega_M$, and the 
density due to dark energy, $\Omega_\Lambda$, the 
expression for the luminosity distance in a 
Friedmann--Lema\^itre--Robertson-Walker (FLRW) Universe without radiation is:
\begin{equation} \label{eq2}
D_L = \frac{c}{H_0}(1+z)| \Omega_K |^{-1/2} sinn \left[ | \Omega_K |^{1/2} \int_{0}^{z} \frac{dx}{E(x)} \right]
,\end{equation}\[
E(z)=\sqrt{\Omega _M(1+z)^3+\Omega _K(1+z)^2+\Omega _\Lambda }
,\]
where $\Omega_K = 1-\Omega_M-\Omega_\Lambda$ and $sinn$ is $sinh$ for $\Omega_K \geq 0$,  
$sin$ for $\Omega_K \leq 0$.
If we give this distance in Mpc, we have the following distance modulus:
\begin{equation} \label{eq3}
\mu_p = 5\log D_L+25
\end{equation}
There is a degeneracy in the values of $H_0$ 
and in the absolute magnitude $M$. To avoid this degeneracy, we set
an equivalent formulation with a single free 
parameter $A$ containing the two previous ones.
The function to be fitted is:
\begin{equation} \label{eq7}
y = 10^{0.2m} = A\,D_{L70}
,\end{equation}
\begin{equation} \label{eq8}
A = \frac{70\ {\rm km\ s^{-1} Mpc^{-1}}}{H_0}10^{0.2M+5}
,\end{equation}
where $D_{L70}$ is the calculated luminosity distance for 
$H_0 = 70$ km s$^{-1}$ Mpc$^{-1}$.
Thus, both $H_0$ and M (absolute magnitude) are included in $A$.

\begin{figure}[ht!]
	\centering
  	\includegraphics[scale=1.0]{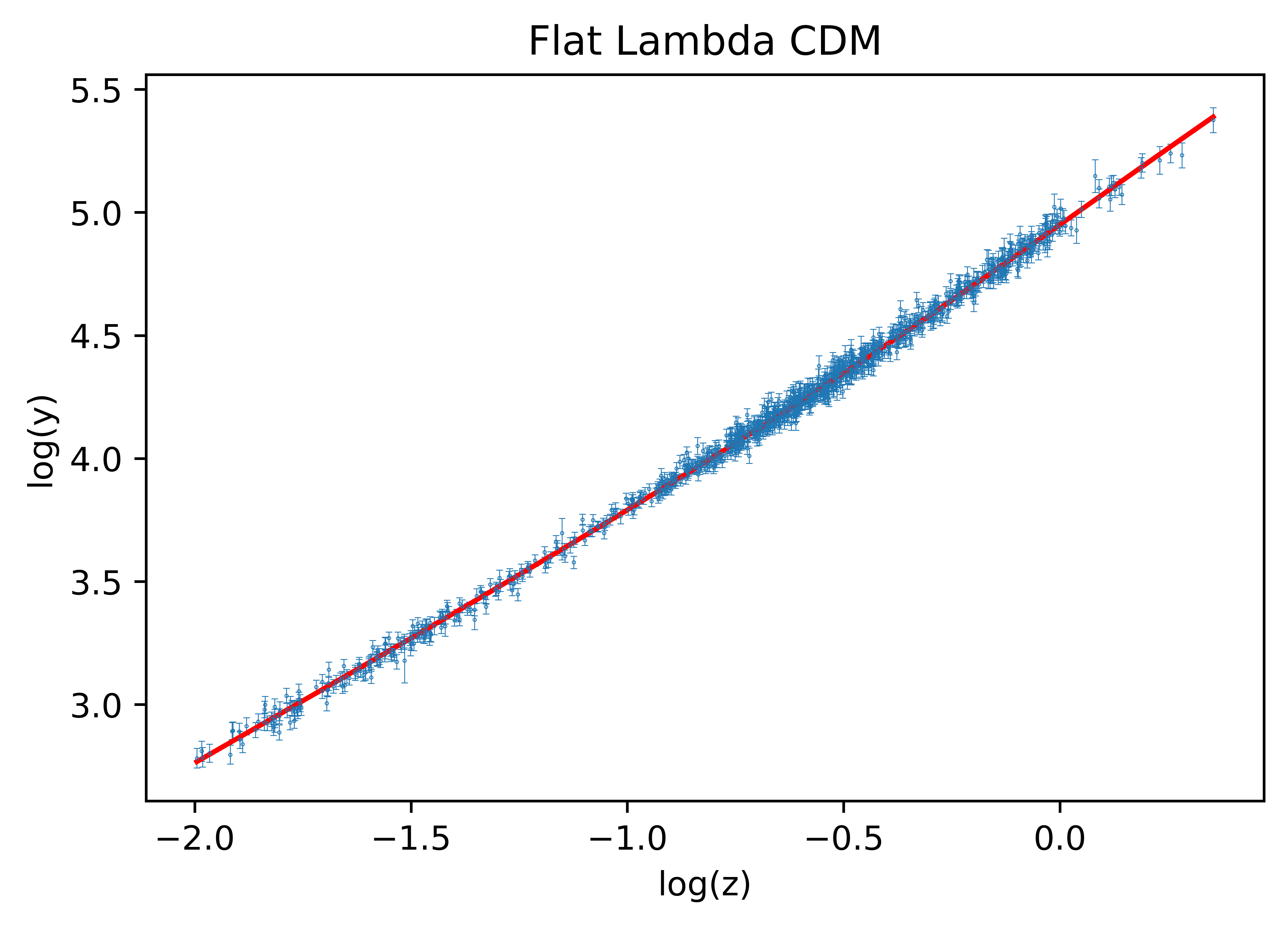}
  	\caption{Fit of Pantheon data with $\Lambda $CDM, $\Omega_M = 0.287$.}
  	\label{fig:flcdm}
\end{figure}

The fit is done through a minimization of $\chi^2$, 
minimizing the differences of $y$ between the data and the theoretical models.
For the standard $\Lambda $CDM model, $\Omega _K=0$, 
leaving $A$ and $\Omega_M$ as free parameters.
Results of the fit are shown in Table \ref{table2}, which are 
compatible with those obtained by Ref.\ \citen{Scolnic2018}. 
We also provide the values of the probabilities $Q$  associated 
with the $\chi^2$ values for the corresponding degrees of freedom 
(=1048 - number of free parameters). 
In Fig.\ \ref{fig:flcdm}, we show the logarithm of the 
quantity $y$ versus the logarithm of $z$. 

The chi-square test already takes into account the difference 
in the number of free parameters by reducing the degrees of 
freedom, and the $Q$ probabilty  depends on these degrees of freedom. 
Information Criterion methods such as AIC, BIC or KIC present 
different estimators with a somewhat higher dependence on 
the number of free parameters, but no major differences arise
 when $N$ is a very large number. They are equivalent 
to a modification of $\chi^2$, $\chi _{\rm mod}^2=\chi^2+u\,p$,
where $p$ is the number of free parameters and $u$ is a 
parameter that depends on the information 
criterion \cite{Lop18}: $u=1$ for the conventional 
minimization of $\chi ^2$; $u=2$ for AIC, $u=3$ for KIC; and 
$u=ln(N)$ for BIC. Given than $N$ is much larger than $u\,p$ 
for any of the methods, the numbers
of $\chi _{\rm mod}^2$ are almost independent of the criterion.

\begin{table}[ht!] 
\caption{}
TABLE 1: Best fits of Pantheon data ($N=1048$ supernovae Ia) 
with different cosmological models, with extinction,  
evolution, or a non-cosmological LSV redshift component. $Q$ 
is the probability associated with $\chi ^2$ with degrees of 
freedom$=N-p$, where $p$ is the number of free parameters.
 
\label{table2}
\centering
\begin{tabular}{|c|c|c|c|}
\hline
Model & Free parameters & $\chi^2$ & $Q$  \\ \hline \hline
$\Lambda $CDM & 
$\begin{array}{ll}
A=13.389\pm 0.041 \\
\Omega_M=0.287\pm 0.012 \\
\Omega _\Lambda =1-\Omega _M 
\end{array}$ 
& 1024.3 & 0.678 \\ \hline
FLRWcurv.$\Lambda $  & 
$\begin{array}{ll}
A=13.320\pm 0.056 \\
\Omega_M=0.355\pm 0.039 \\
\Omega _\Lambda =0.835\pm 0.067 
\end{array}$ 
& 1021.1 & 0.696 \\ \hline
FLRWcurv$\Lambda =0$ & 
$\begin{array}{ll}
A=13.909\pm 0.039 \\
\Omega_M=0 \pm 0.024
\end{array}$
& 1147.1 & 0.015 \\ \hline
EdS & $A=14.832\pm 0.043$ & 2395.4 & 0 \\ \hline
QSSC & 
$\begin{array}{ll}
A=14.444\pm 0.067 \\ 
\Omega _M=1.439\pm 0.068 \\
\Omega _\Lambda =0\pm 0.042 \\
\Omega _c=1-\Omega _M-\Omega _\Lambda
\end{array}$ 
& 1635.6 & 0 \\ \hline
$R_h=ct$ & $A=14.091\pm 0.030$ & 1296.3 & 0 \\ \hline
Milne & $A=13.909\pm 0.029$  & 1147.1 & 0.016 \\ \hline
Static.lin.Hub. & $A=14.047\pm 0.029$  & 1239.1 & 0 \\ \hline
Static tired light & $A=15.559\pm 0.061$ & 4374.1 & 0 \\ \hline
St.tir.l. Compton & $A=12.493\pm 0.033$ & 2021.0 & 0 \\ \hline
St.tir.l.tim.dil. & $A=14.091\pm 0.030$ & 1296.3 & 0 \\ \hline \hline
EdS extinction & 
$\begin{array}{ll}
A=13.487\pm 0.045 \\ 
a_V=(1.156\pm 0.031)\times 10^{-4}\ {\rm Mpc}^{-1}
\end{array}$ 
& 1050.8 & 0.453 \\ \hline
St.lin.Hub.ext. & 
$\begin{array}{ll}
A=13.566\pm 0.045 \\ 
a_V=(0.403\pm 0.031)\times 10^{-4}\ {\rm Mpc}^{-1}
\end{array}$ 
& 1065.2 & 0.333 \\ \hline
St.tir.l.ext. & 
$\begin{array}{ll}
A=12.959\pm 0.051 \\ 
a_V=(2.775\pm 0.048)\times 10^{-4}\ {\rm Mpc}^{-1}
\end{array}$
& 1072.9 & 0.275 \\ \hline \hline
$\Lambda $CDM evol. & 
$\begin{array}{ll}
A=13.257\pm 0.084 \\
\Omega_M=1\pm 1.406 \\
\Omega _\Lambda =1-\Omega _M\\
\alpha =-0.102\pm 0.178\ {\rm Gyr}^{-1} 
\end{array}$ 
& 1020.4 & 0.701 \\ \hline
FLRWcurv.$\Lambda $,evol.  & 
$\begin{array}{ll}
A=13.268\pm 0.084 \\
\Omega_M=0.957\pm 0.162 \\
\Omega _\Lambda =0 \\ 
\alpha =-0.099\pm 0.066\ {\rm Gyr}^{-1}
\end{array}$ 
& 1020.3 & 0.702 \\ \hline
EdS evol. & 
$\begin{array}{ll}
A=13.257\pm 0.048 \\
\alpha =-0.102 \pm 0.003\ {\rm Gyr}^{-1}
\end{array}$ 
& 1020.4 & 0.709 \\ \hline \hline
\end{tabular}
\end{table}

\begin{table}[ht!] 
Cont. Table \ref{table2}.
\centering
\begin{tabular}{|c|c|c|c|}
\hline
Model & Free parameters & $\chi^2$ & $Q$    \\ \hline \hline
EdS+LSV1 & 
$\begin{array}{ll}
A=31.445\pm 0.119 \\
K_1=(-2.046\pm 0.005)\times 10^{-4}\ {\rm Mpc}^{-1}
\end{array}$ 
& 1490.7 & 0 \\ \hline
EdS+LSV2/3 & 
$\begin{array}{ll}
A=38.263\pm 0.212 \\
K_2=(-2.113\pm 0.003)\times 10^{-4}\ {\rm Mpc}^{-1}
\end{array}$ 
 & 1194.8 & 0.0008 \\ \hline
EdS+LSV4 & 
$\begin{array}{ll}
A=29.264\pm 0.103 \\
K_4=(-2.012\pm 0.006)\times 10^{-4}\ {\rm Mpc}^{-1}
\end{array}$ 
& 1482.1 & 0 \\ \hline 
$\begin{array}{ll}
{\rm FLRWcurv.}\Lambda =0\\
+LSV1
\end{array}$
 & 
$\begin{array}{ll}
A=9.479\pm 0.712 \\
\Omega_M=0.090\pm 0.015 \\
K_1=(0.720\pm 0.060)\times 10^{-4}\ {\rm Mpc}^{-1}
\end{array}$ 
& 1091.1 &  0.157   \\ \hline 
$\begin{array}{ll}
{\rm FLRWcurv.}\Lambda =0\\
+LSV2/3
\end{array}$
 & 
$\begin{array}{ll}
A=40.563\pm 0.192 \\
\Omega_M=0.190\pm 0.010 \\
K_2=(-2.220\pm 0.010)\times 10^{-4}\ {\rm Mpc}^{-1}
\end{array}$ 
& 1082.0 &  0.208   \\ \hline
$\begin{array}{ll}
{\rm FLRWcurv.}\Lambda =0\\
+LSV4
\end{array}$
 & 
$\begin{array}{ll}
A=3.041\pm 1.040 \\
\Omega_M=0.260\pm 0.030 \\
K_4=(1.800\pm 0.018)\times 10^{-4}\ {\rm Mpc}^{-1}
\end{array}$ 
& 1019.9 &  0.705   \\ \hline \hline
\end{tabular}
\end{table}

\section{Alternative cosmologies}

In this section we will fit Pantheon data for alternative cosmologies different 
from the standard $\Lambda $CDM model \cite{Corredoira2010,Corredoira2016}. Results in Table \ref{table2}.

\begin{description}
\item[FLRW with curvature and dark energy:]
We repeat the fit of the standard model, but for a 
model in which we do not impose a flat universe.
We obtain a slightly better fit and a significantly greater value for $\Omega_M$.
We obtain as best fit a curvature parameter of $\Omega _K=-0.19$.
However, although the fit to the supernova data is 
better, other observations show that the universe has 
effectively zero curvature, as shown, for example, by 
 Cosmic Microwave Background Radiation (CMBR) analyses \cite{Planck2014}.

\item[FLRW with curvature without dark energy:]
We now consider an FLRW cosmology without dark energy
 and satisfying $\Omega_K = 1 - \Omega_M$, 
leaving both $A$ and $\Omega_M$ as free parameters.
In this way, we obtain a negative-curvature cosmology (flat if $\Omega_M = 1$).
In the fit we obtain a value of practically zero  got $\Omega_M$. 
The fit is worse than in the case of the $\Lambda 
$CDM and FLRW model  without curvature and dark energy. 
This is equivalent to the Milne Universe that we will comment later.
The low value of probability $Q$ almost completely rules out this model.

\item[Einstein-de Sitter:]
The Einstein-de Sitter (EdS) model is equivalent to an 
FLRW cosmology with $\Omega_M = 1$ and $\Omega_\Lambda = 0$.
We can use the same fitting method but imposing those 
values, with a single free parameter ($A$).
As expected, such a restricted model obtains a very poor 
fit and is completely discarded.
As we will see later, if we add certain factors to this 
model (evolution, extinction, or partially
non-cosmological redshifts) we can obtain much better fits.

\item[Quasi-steady state:]
The Steady State model establishes a universe that, in 
addition to the cosmological principle (homogeneity and 
spatial isotropy), is homogeneous in time.
To compensate for the expansion of the universe, there 
is a field that continuously creates matter to keep the 
density constant.
This model presented several problems and was later
 modified to become the quasi-steady state (QSSC) model.
In addition to the matter creation field, we have an 
expansion with an oscillatory term.
In this model, calling $\Omega_c$ the matter creation field we have:
\begin{equation} \label{eq9}
D_L(z) = \frac{c}{H_0}(1+z)\int_{0}^{z}\frac{dz}{\sqrt{\Omega_c (1+z)^4 + \Omega_m (1+z)^3 + \Omega_\Lambda)}}
\end{equation}
We see that the term $\Omega_c$ behaves like a radiation 
term in FLRW, which also goes as $(1+z)^4$.
A fundamental problem with QSSC is that no galaxies 
should be observed at $z > 6$, but galaxies have been 
observed with redshifts as high as $z = 8.6$ \cite{Lehnert2010}. With 
the Pantheon data and without taking into account other 
factors, we obtain a fit that completely rules out this model.
There have been previous attempts to fit supernova Ia
 data with this model \cite{Banerjee2007} with apparently 
very good results, but in that paper they also take dust 
extinction into account.

\item[$\mathbf{R_h=ct}$:] This is an FLRW model in which 
we have an equation of state, $\rho + 3p = 0$, thus leading 
to expansion, $R_h = c\,t$.
The luminosity distance is:
\begin{equation} \label{eq10}
D_L(z) = \frac{c}{H_0}(1+z)ln(1+z).
\end{equation}
The fit, being better than the QSSC and the Einstein--de Sitter models, is 
still ruled out by the Pantheon data.
Other authors \cite{Melia2012} have stated that this model 
fits the supernovae diagram, but only after re-evaluating the 
calibration of their luminosities and putting their absolute 
magnitude as a function of several adjustable parameters 
instead of being constant.

\item[Milne Universe:]
This model is a special case of the FLRW metric in which we 
consider zero density, pressure, and cosmological constant.
This results in a linear time dependence of the scale factor.
We have the following expression for the luminosity distance:
\begin{equation} \label{eq11}
D_L(z) = \frac{c}{H_0}(1+z)sinh[ln(1+z)].
\end{equation}
Despite being a model with very restricted parameters, the 
fit is better than that of the other models considered. However, 
Milne's universe is unable to explain other contrasting facts 
of cosmology and such as the CMBR, the abundance of light elements.

\item[Static Euclidean with linear Hubble-Lema\^itre law:]
In this case, we consider a static universe in which we have 
a redshift term due to energy loss without expansion (no time
 dilation), and the linear Hubble--Lema\^itre law $c\,z=H_0 D$
 is maintained even at high redshift. The luminosity distance is:
\begin{equation} \label{eq12}
D_L(z) = \frac{c}{H_0}\sqrt{(1+z)}z.
\end{equation}
The factor $\sqrt{(1+z)}$ stems from the loss of energy of
 photons due to  non-cosmological redshift.
We obtain a setting that rules it out. Moreover, although 
it is not the subject of this work, a simple static model 
has many other problems.

\item[Static Euclidean with tired light:]
This variation considers that photons lose energy in their
 path by some interaction, and that this energy loss is proportional to the length traveled,
$\frac{dE}{dr} = - \frac{H_0}{c}E$. This modifies the luminous distance as follows:
\begin{equation} \label{eq13}
D_L(z) = \frac{c}{H_0}\sqrt{(1+z)}ln(1+z)
.\end{equation}
In view of the results, this consideration of energy loss 
is totally incompatible with the Pantheon data and greatly
 worsens the fit of the simplest linear model.

\item[Static Euclidean with tired light-plasma/Compton:]
First, we assume a Compton scattering factor, modifying the expression to become:
\begin{equation} \label{eq14}
D_L(z) = \frac{c}{H_0}(1+z)^{3/2}ln(1+z).
\end{equation}
Although it represents an improvement in the fit with respect 
to the tired light model, it is still far from the simple 
linear model, which had already been discarded.

\item[Static Euclidean with tired light-plasma/time dilation:]
Let us try another tired photon model with a scattering that 
gives rise to a time dilation that broadens the light curves 
of supernovae with redshift:
\begin{equation} \label{eq15}
D_L(z) = \frac{c}{H_0}(1+z)ln(1+z).
\end{equation}
In this case we have a much better fit that is still, however, ruled out by the data.

\end{description}

\section{Including extinction}

A progressive flux reduction with redshift is seen in the 
supernova data. This can be explained by a universe with a 
positive cosmological constant (dark energy).
Another explanation is the presence of dust particles in the 
medium that passes through the light. This dust absorbs 
light in the optical and re-emits it in the far infrared.
Riess et al.\ state that the effect of 
extinction is too small to be taken into account in the 
analysis. Still, with more recent data, with more supernovae 
measured with greater precision, it is important to 
reanalyze this effect to see whether it is really minimal, 
or whether  there is simply
no need for a dark energy term. Ref.\ \citen{Goobar2002} 
consider three types of extinction:

\begin{description}

\item[Large gray intergalactic dust grains:] 
Following a model of Ref.\ \citen{Draine1984}, large grains 
are considered because the smaller ones have been destroyed. 
Ref.\ \citen{Goobar2002} perform a Monte Carlo simulation 
to obtain an adjustment of the cosmological parameters while 
taking into account this extinction factor (whose values 
depend, in turn, on the cosmological model itself).

\item[Dust in galaxies along the line of sight:] 
In this case, they perform several ray-tracing Monte 
Carlo simulations to estimate the probability, as a function 
of $z$, of a ray passing close to the center of a galaxy.

Again, without going into much detail, they find that 
the probability of a supernova at $z \sim 1$ being 
obscured by plus 0.02 mag is only 0.33\%. They claim 
that this value is highly dependent on the dust density 
normalization itself. On the other hand, a galaxy acts 
as a gravitational lens that would have the opposite 
effect of extinction dimming.

\item[Extinction in the host galaxy:] 
Since an SN Ia occurs in both early- and late-type 
galaxies, obscuration due to dust from the host galaxy 
would have to be considered in the late-type case. By 
modeling the galaxy and estimating at 1/8 the probability 
of the supernova occurring in the bulge rather than
in the disk, the simulations performed conclude that 
about 2600 out of 10000 sources woulf be obscured by more 
than 0.02 mag by dust from the host galaxy. Nonetheless, 
attention to new analyses that show that Type Ia 
Supernova brightness correlates with host galaxy dust content
\cite{Mel22}, which might be interpreted as an indication that dust
in host galaxies is more important than we thought.
\end{description}

Ref.\ \citen{Goobar2002} conclude that the overall effect 
of extinction would be very small, since only  1\% of 
the supernovae at $z=1$ would be obscured by more than 
0.02 mag (the detection limit of the SNAP probe).

\subsection{Fits including extinction}

\begin{figure}[ht!]
\centering
\includegraphics[scale=.6]{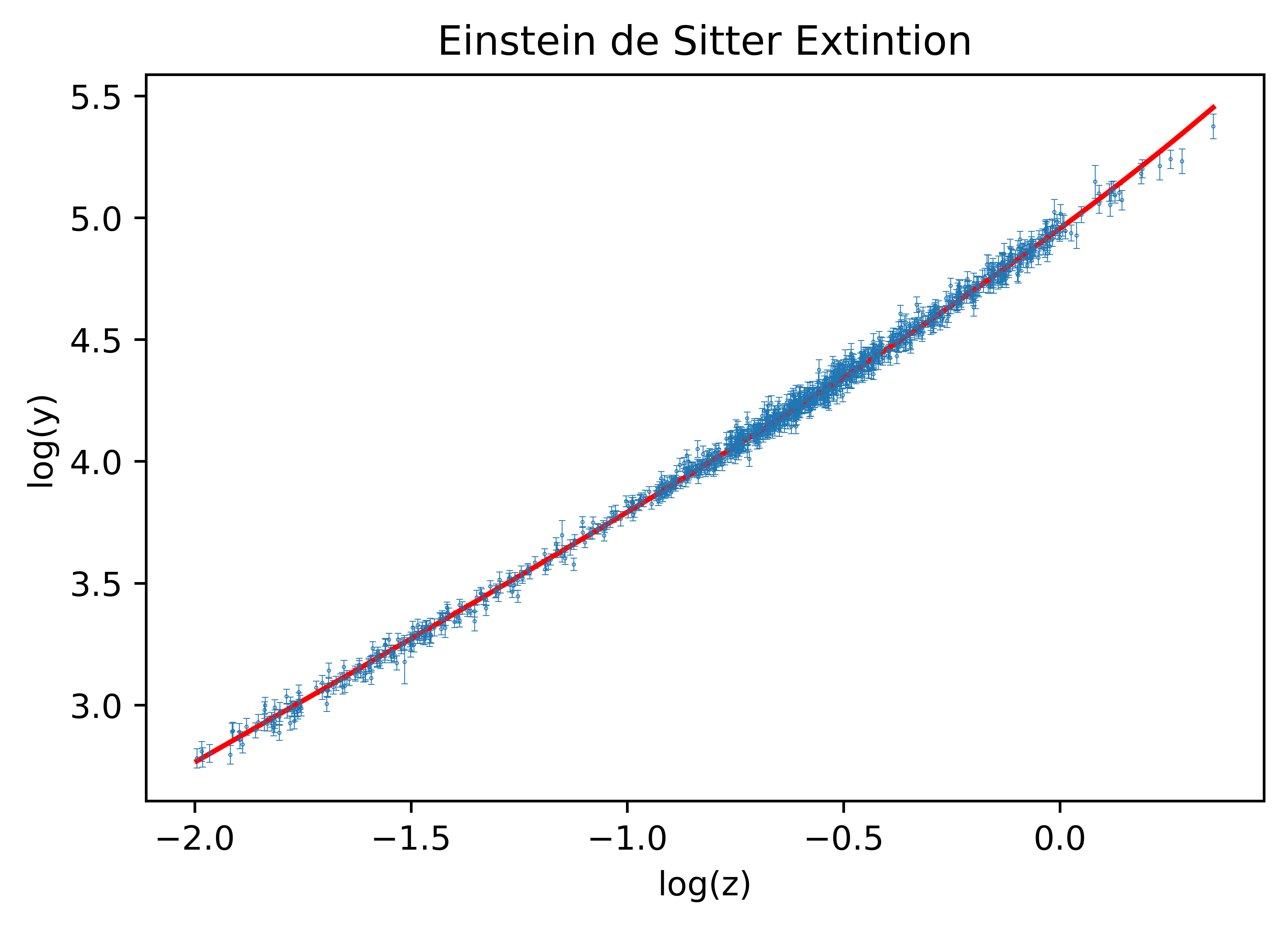}\\
\includegraphics[scale=.6]{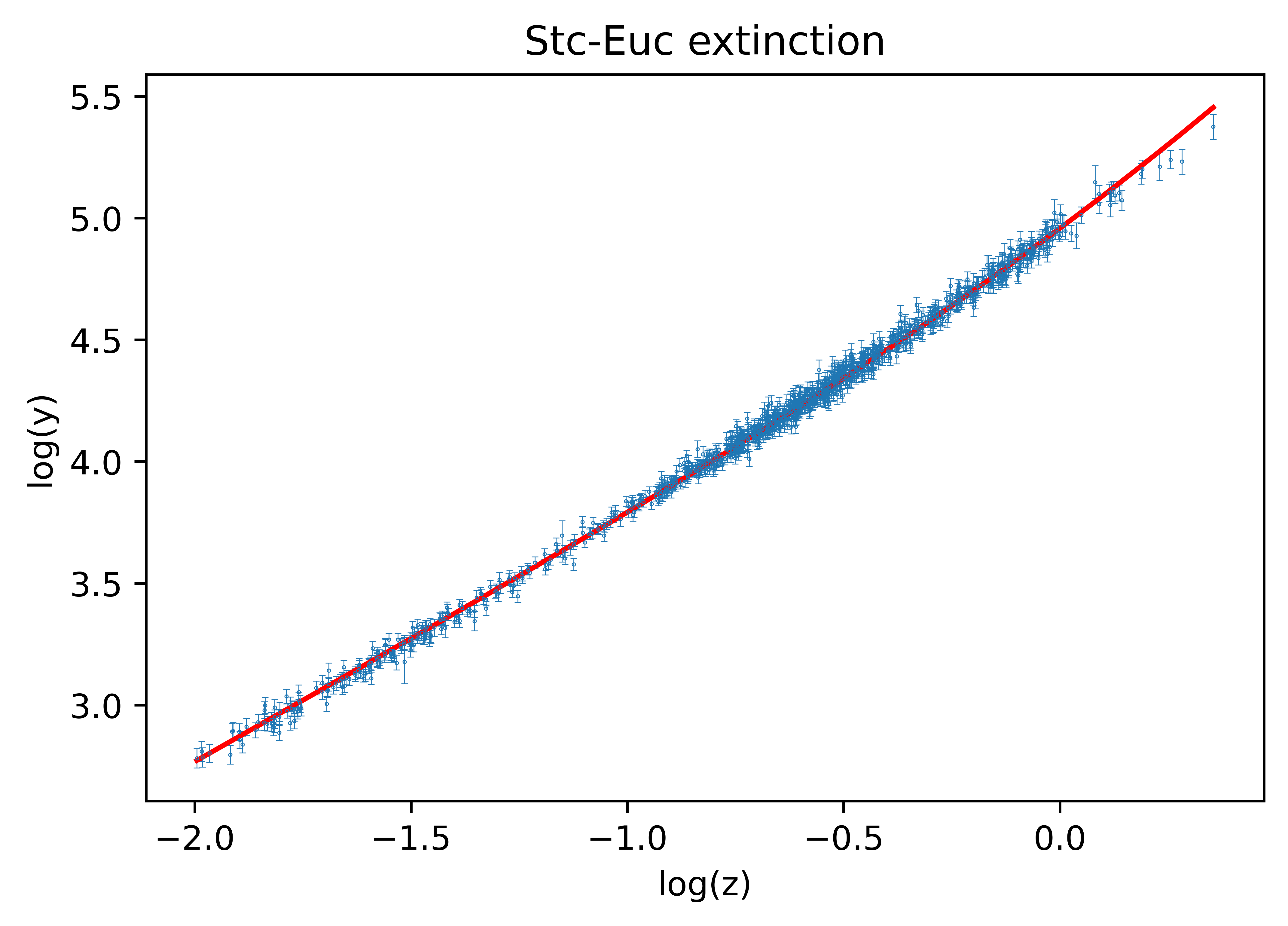}\\
\includegraphics[scale=.6]{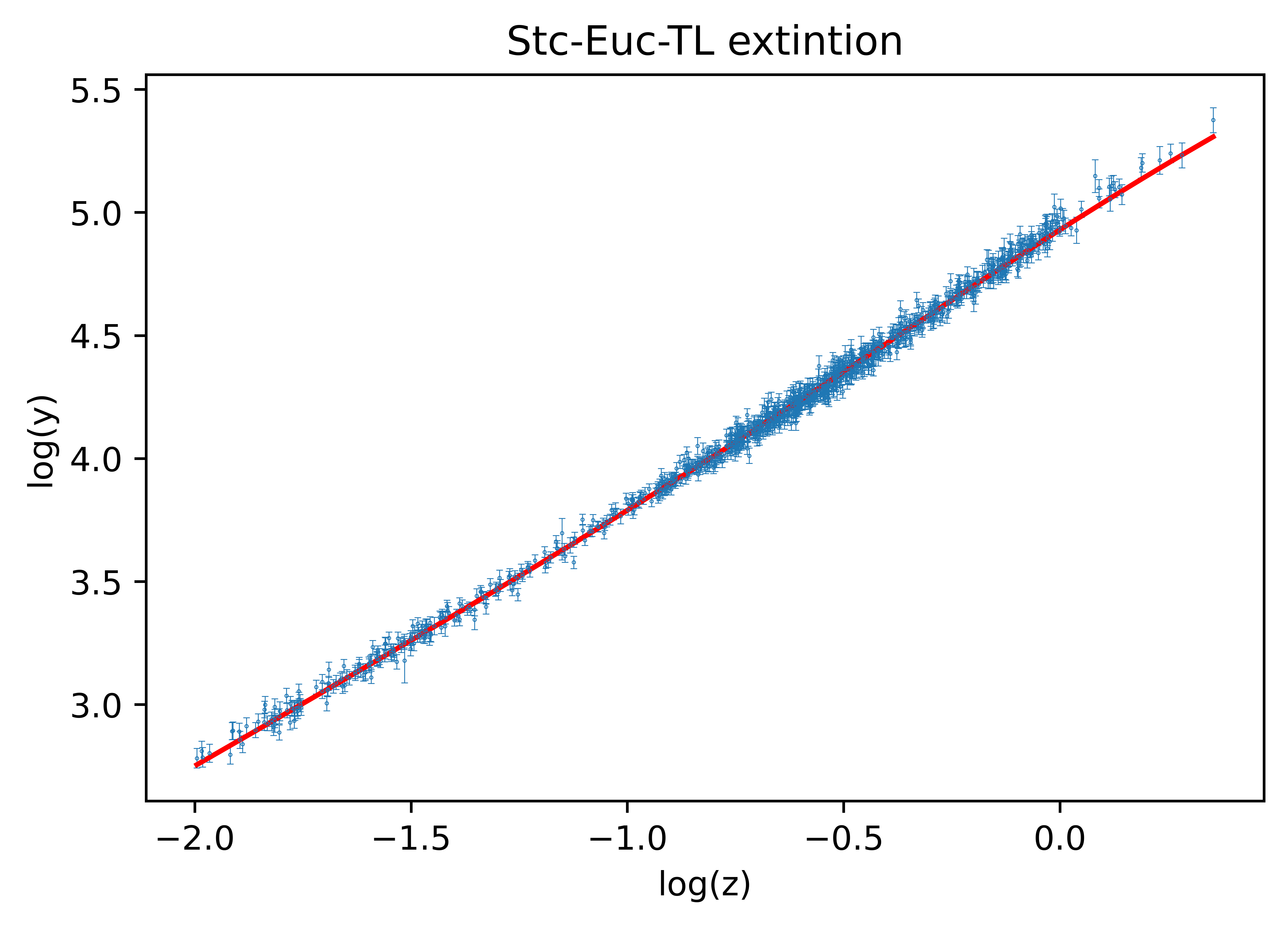}
\caption{Fit of Pantheon data with Einstein de--Sitter, 
Static Euclidean with linear Hubble law, Static Euclidean 
with tired light cosmological models including 
$a_V = 1.156\times 10^{-4},\ 0.403\times 10^{-4}, 2.775\times 10^{-4}$ Mpc$^{-1}$ respectively.}
\label{fig:ext}
\end{figure}

Introducing an extinction term will make very distant 
galaxies appear less luminous. This effect can counteract 
that of dark energy and thus provide a good fit \cite{Corredoira2010}.
Assuming a constant comoving dust density term, $\kappa$ 
being the absorption coefficient per unit mass, we have 
the following expression for luminosity:

\begin{equation} \label{eq16}
L_{V,rest} = 4\pi F_{V,rest}D^2_L e^{\rho_{dust}\int^{d}_0 dr\,\kappa\left[\lambda_V\frac{1+z(d)}{1+z(r)}\right]}
,\end{equation}
where the comoving distance $d$ in the corresponding 
cosmology associated with the redshift of the supernova. 
We assume a absorption coefficient with a wavelength dependence,
\begin{equation} \label{eq17}
\kappa(\lambda) = \kappa(\lambda_V)\left(\frac{\lambda}{\lambda_V}\right)^{-\beta}
,\end{equation} 
and  adopt $\beta = 2$ \cite{Corredoira2010}.
With this we obtain:
\begin{equation} \label{eq18}
L_{V,rest} = 4\pi F_{V,rest}D^2_L e^{\frac{c\,a_V}{H_0(\beta +m)}[(1+z)^m-(1+z)^{-\beta}]}
,\end{equation}
$a_V \equiv \kappa(\lambda_V)\rho_{dust}$ being the 
absorption in V per unit length.
Together with $A$, we shall fit the parameter $a_V$. 
Regarding the other parameters, $m=1$ for the 
Einstein de--Sitter and Linear Static models; and $m=0$ for the 
static model with tired light. See the best fits in 
Fig. \ref{fig:ext}. The attenuations 
$a_V = 0.4-2.8\times 10^{-4}$ Mpc$^{-1}$ obtained  are within the
possible range of values. Assuming $\kappa (\lambda_V)\sim 10^5$
cm$^2$/gr \cite{Wic96}, the value for the dust density 
necessary to produce such an extinction
would be $\rho_{dust}\sim 10^{-34}-10^{-35}$ g/cm$^3$,
which is within the range of possible values. 
Ref.\ \citen{Ino03} allow values as
high as $\rho_{dust}\sim 10^{-33}$ g/cm$^3$ for the high 
$z$ IGM in the standard concordance
cosmology. For comparison, the average baryonic density 
of the Universe (taking
$\Omega _b = 0.042$) is $\rho_{b}= 3.9 \times 10^{-31}$ 
g/cm$^3$, so this would mean that IGM dust
constitutes 0.025--0.25\% of the total baryonic matter 
(reasonable amounts). Results are shown in Table \ref{table2}.

\begin{description}

\item[Einstein-de Sitter with extinction]
We see a substantial improvement in the fit over the model without extinction.
If we consider only the Pantheon data, this would be a model to consider.

\item[Static Euclidean with linear Hubble law with extinction:]
We see that the fit is also considerably better when an extinction term is included.

\item[Static Euclidean with tired light with extinction:]
Despite being a worse fit than the previous two, 
compared to the model without extinction, the improvement 
is great indeed (going from $\chi ^2=4374.1$ to 1072.9).

\end{description}

\section{Including evolution}

One of the assumptions made in the work of Riess et al.\  
was that the shape of supernova light curves has negligible 
variation with redshift, so that well-defined models of 
nearby supernovae
 could be applied to distant supernovae.
With more recent data, reaching higher redshift, 
perhaps this effect can no longer be neglected completely.
At higher redshift we have more massive progenitor stars 
with lower metallicity. This influences the composition 
of their degenerate CO core. This variation may have implications for ignition 
due to accretion, so that the supernova light curve 
varies, having a difference of 0.2 mag for the luminosity 
maximum, comparable to the brightness variation indicating dark energy
 \cite{Dominguez1999}. Or there may possibly be evolution 
of the luminosity if there is an evolution of the physical 
constants \cite{Gup22}.

As a matter of fact, there is a systematic difference of 
$\sim 0.14$ mag between supernovae whose host galaxies are of 
very early and very late type \cite{Kang2020}.
In addition, there is a correlation between the mass of the 
host galaxy and the brightness of the SNe Ia, suggesting 
that in less massive galaxies we have $\sim 0.1$ mag lower brightness
than in massive galaxies.
There is also a correlation indicating that in high star 
formation environments, SNe Ia are less bright than in 
more passive environments. 
All of these correlations actually indicate a relationship 
between supernova brightness, and the age and metallicity of the progenitor.
We have a brightness variation due to evolution that is 
comparable to the margin indicated by a positive cosmological constant ($\sim 20$\%).
This has led several researchers to point out that an 
evolution of SNe Ia luminosity can fit their Hubble--Lema\^itre 
diagram without dark energy \cite{Kang2020,Dre00,Wri02,Tut19,Lee2020,Lee21}.

\subsection{Fits including evolution} 

In our analysis, we use a simple expression to model 
cosmic time-dependent evolution and a parameter $\alpha$. 
We fit cosmological models without dark energy. 
We assume that the absolute magnitude of supernovae varies 
over cosmological time,  using the simple evolution equation
\begin{equation} \label{eq22}
M = M_0-\alpha[t(0)-t(z)],
\end{equation}
where $M_0$ is the absolute magnitude without taking
evolution into account and $t(z)$ is the age of the universe at a given $z$ 
[$t(0)$ for $z=0$].
We  calculate that age for each given $z$  and incorporate 
the new $M$ into the formula for the $A$ parameter fit:
\begin{equation} \label{eq23}
A_{\rm evol} = A \times 10^{-0.2 \alpha [t(0)-t(z)]} 
,\end{equation}
where
\begin{equation} \label{eq4}
t(z) = \frac{1}{H_0} \int_{z}^{\infty} (1+x)^{-1} E(x) dx
.\end{equation}

Results in Table \ref{table2}.
\begin{description}

\item[$\Lambda $CDM with evolution:]
We repeat the setting for the standard model but with a 
magnitude that takes into account the evolution with redshift.
 We get a small improvement in the value of $\chi^2$. But
 the most striking thing is that we obtain a value of 
$\Omega_M$ of practically 1, although with a very large error.
 That value would produce (for a flat metric) 
$\Omega_\Lambda \sim 0$. That is, introducing an evolutionary 
term allows us to fit the data without the need for dark energy.
Refs.\ \citen{Lee2020,Lee21} obtained a correlation 
magnitude--age of galaxies (early type galaxies having a passive
evolution, which is supposed to trace the age of the universe)
with lower $\alpha $ for $\Omega _m=0.27$, $\Omega _\Lambda =0$, and for 
a lower reshift sample: $\alpha $ between -0.06 and -0.04 Gyr$^{-1}$.

\item[Friedmann with curvature and dark energy and with evolution:]
We do the same but in this case we do not impose a null curvature.
We impose a starting value of zero for dark energy (since 
we obtained zero for the $\Lambda $CDM case without curvature) 
to see if evolution can fit the data well without taking it into consideration.
By adjusting one parameter less, we have a lower uncertainty 
for the value of $\Omega_M$. This means  that the 
value is very close to unity, but somewhat smaller. Since $\Omega_\Lambda = 0$, 
that implies a positive curvature of $\Omega _K=0.043$.
Regarding the fit itself, we get a slightly better value. 
That is, with the Pantheon data, we get better fits by 
introducing evolution (and also with a very simple model) 
than when considering the dark energy term alone.

\item[Einstein--de Sitter with evolution:]
Finally we fit to an Einstein--de Sitter model with evolution, 
which is equivalent to a $\Lambda CDM$ model but imposing 
$\Omega_M = 1$ and $\Omega_\Lambda = 0$.
We obtain a very good fit. This is the model that gives a better 
value for the $Q$ probability. In addition, we have smaller 
uncertainties by fitting only two parameters ($A$ and $\alpha $).
\end{description}

\begin{figure}[ht!]
	\centering
  	\includegraphics[scale=1.0]{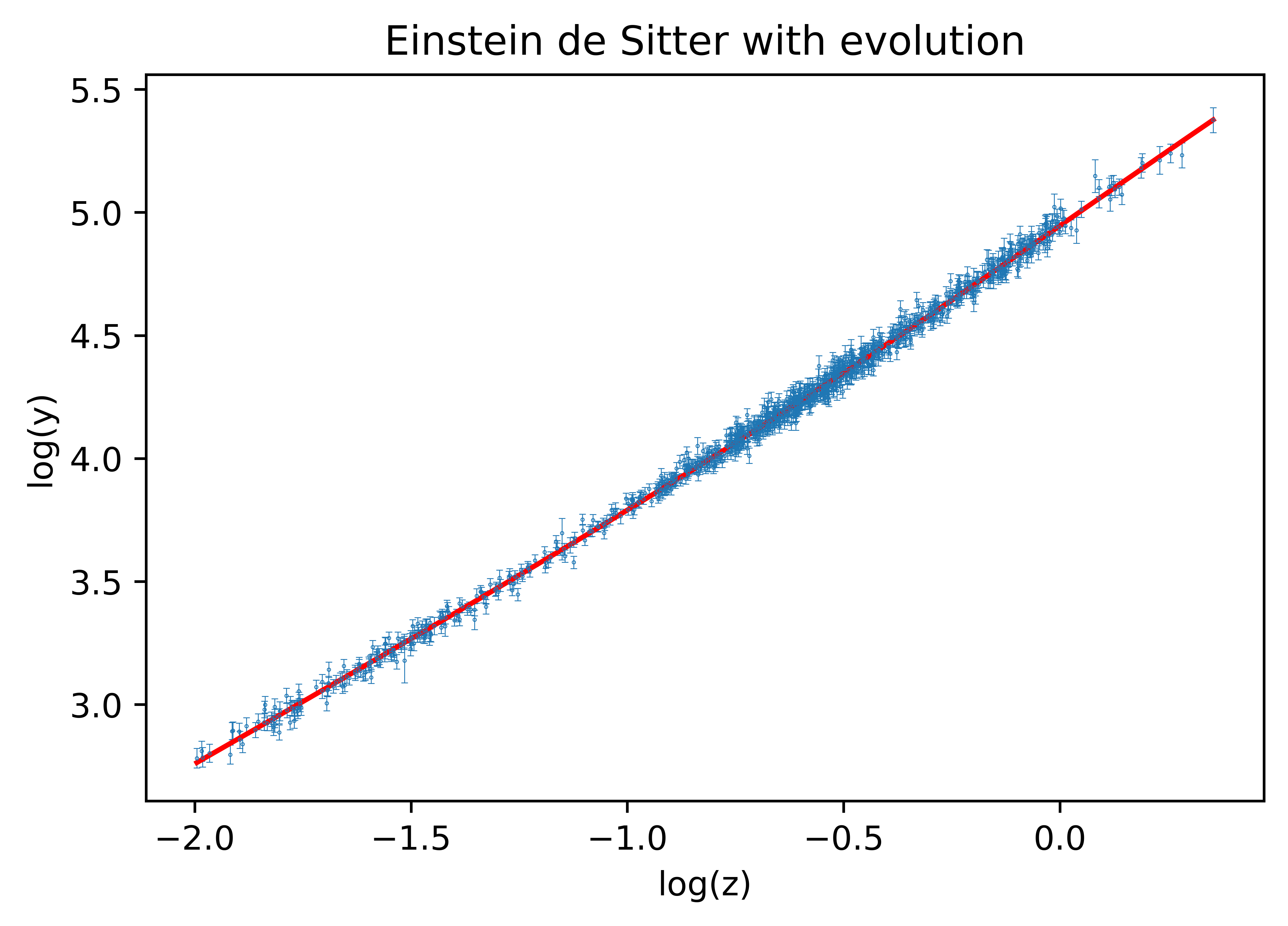}
  	\caption{Einstein--de Sitter with evolution:
 $\Omega_M = 1$, $\Omega_\Lambda = 0$, $\alpha =-0.102$ Gyr$^{-1}$.}
  	\label{fig:eds_evol}
\end{figure}

We see, then, that with the Pantheon data a model with evolution 
is just as valid (or better) than one with dark energy. 
Moreover, when trying to fit with both terms, the best value 
of $Q$ is for the one that gives  $\Omega_\Lambda \sim 0$. See 
the best fit in Figure \ref{fig:eds_evol}.

As said by Ref.\ \citen{Dre00},
the result is that cosmological models and evolution are highly 
degenerate with one another, so that the incorporation of evolution into
even very simple models renders it virtually impossible to pin down the 
values of $\Omega _M$ and $\Omega _\Lambda $.

\section{Partially non-cosmological redshifts}

If we add a partially non-cosmological redshift or blueshift 
we will obtain supernova distances  that are different from 
those obtained in equation (\ref{eq2}).
The non-cosmological redshift component may be
due to the non-conservation of the energy-momentum tensor of 
a photon propagating through electromagnetic fields 
[Lorentz--Poincar\'e Symmetry Violation (LSV)] \cite{Spallicci2021} 
or Mach effects that relate tired light
with the mass of the Universe \cite{Gup18,Gup19}, or other non-standard effects.
 The Hubble-Lema\^itre diagram could be fitted without the need for 
dark energy. We now study four simple non-cosmological redshift models.

\subsection{Fits including non-cosmological redshifts}

We consider models in which the total redshift is the 
sum of the redshift due to expansion plus a non-cosmological term.

The excessive dimming of very distant galaxies can be 
attributed to this combined redshift without the need to 
include a dark energy term.
For a SN Ia we have the relation:
\begin{equation} \label{eq24}
m = M + 5\,\log_{10}[d_L(z_c)({\rm Mpc})]+25
\end{equation}
\begin{equation} \label{eq25}
z_c = \frac{1+z}{1+z_{\rm LSV}}-1,
\end{equation}
Where $z_c$ is the cosmological redshift and $z_{\rm LSV}$ is the 
non-cosmological redshift, which depends on the model. We 
have four different models, as seen in Table \ref{Tab:noncosmz}.

\begin{table}[ht!]
\caption{}
TABLE 2: Four different models with  non-cosmological redshifts.
\label{Tab:noncosmz}
\centering
\begin{tabular}{|c|c|c|c|c|}\hline
Type & 1 & 2 & 3 & 4 \\ \hline
$d_\nu$ & $k_1\nu dr$ & $k_2\nu_e dr$ & $k_3 dr$ & $k_4\nu_0 dr$ \\ \hline
$\nu_0$ & $\nu_e e^{k_1r}$ & $\nu_e(1+k_2r)$ & $\nu_e+k_3r$ & $\frac{\nu_e}{1-k_4r}$ \\ \hline
$z_{LSV}$ & $e^{-k_1r}-1$ & $-\frac{k_2r}{1+k_2r}$ & $-\frac{k_3r}{\nu_e+k_3r}$ & $-k_4r$ \\ \hline
$r_{LSV}$ & $-\frac{ln(1+z_{LSV1})}{k_1}$ & $-\frac{z_{LSV2}}{k_2(1+z_{LSV2})}$ & $-\frac{\nu_ez_{LSV3}}{k_3(1+z_{LSV3})}$ & $-\frac{z_{LSV4}}{k_4}$ \\ \hline
\end{tabular}
\end{table}

In model 1, the variation in frequency is proportional 
to the instantaneous frequency and the distance; in model 
2, to the emitted frequency ($\nu _{\rm e}$) and the distance; 
in model 3 only to the distance; and in model 4, to the 
observed frequency ($\nu_0$). We see that model 3 is 
equivalent to model 2 if we make $k_3 = \nu_{\rm e} k_2$. Since 
the Pantheon data are all in the B-band, we assume the value 
$\nu_e = 6.74\times 10^{14}$ s$^{-1}$.
We need an iterative function to obtain the values of $z_c$, $r$, 
and $z_{\rm LSV}$, since they depend on each other. For this we 
will use the equations (\ref{eq24}), (\ref{eq25}), and the 
relations in Table \ref{Tab:noncosmz}.
\begin{equation} \label{eq26}
r(z_c) = c[t(0)-t(z_c)] 
,\end{equation}
where $t(z)$ is the age of the universe at redshift $z$, 
given by Equation (\ref{eq4}). 
Positive values for $k$ indicate a blueshift with a small, 
or even decreasing, change in the amplitude $A$. 
This non-cosmological blueshift increases the magnitude of 
the SNe Ia at high redshift.
With negative values, we obtain a non-cosmological redshift 
that increases the value of $A$ 
and a reduction in photon frequency with distance traveled.

Results of best fits are shown in Table \ref{table2}.

\begin{description}
\item[Einstein--de Sitter + LSV non-cosmological redshift:]
With model 1, by including the non-cosmological redshift we 
obtain a better fit than the original one, although still far 
from that obtained by including extinction or evolution. 
In view of the results, the Pantheon data would rule out this 
model. The negative value of $k_1$ indicates that we have redshift.
Model 2 and 3 produce a
better fit than model 1, but are also practically ruled out ($Q = $0.0008).
Model 4 gives a result very similar to that of model 1, 
so it is also discarded.
Similar results were obtained with the Pantheon sample 
by Ref.\ \citen{Spa22}. 
 
\item[FLRW with curvature, $\Omega _\Lambda =0$ + LSV 
non-cos\-mo\-lo\-gi\-cal redshift:]
Good fits for models 1--3, and a very good fit for model 4, 
with a quite remarkable best free parameter 
$\Omega _M=0.26\pm 0.03$ (hence, $\Omega _K=0.74\pm 0.03$). 
However, model 4 gives an amplitude
$A=3.0\pm 1.0$, which is $>4$ lower than the value for $A$ with $\Lambda $CDM.
According to Eq.\ (\ref{eq8}), this would mean that the 
absolute magnitude of SNe Ia is 3 magnitudes 
brighter than the local calibrated ones, which is an 
absurd result. For models 1--3 the discrepancy
among the $A$ values is also high, much larger than the inaccuraccies 
in the measurement of $M$ for local SNe Ia.
Moreover, this cosmological model with such high curvature 
is also discarded by many other kinds of observational
cosmological data. Nonetheless, the mathematical fit, 
especially with model 4, for whatever reason,
is remarkably good. Results with the Pantheon sample of 
Ref. \citen{Spa22} were similar, although
with an unclear dependence of $A$ given that the type of fit
 was different.

\end{description}

\section{Conclusions} 

The original 1998--1999 work of Refs. \citen{Riess1998,Per98} 
relied on data from few supernovae with a small range of 
redshifts. Even so, those papers served to establish the 
dark energy model with an accelerating expanding universe 
($\Lambda $CDM). It is cautiously stressed that, owing to 
the paucity of data in the sample and the uncertainties 
involved, such a model cannot be taken as established beyond doubt.
In subsequent years, with a larger amount of data and coverage of 
a wide range of redshifts, the fits have been 
repeated. Many of these papers confirmed the curveless 
dark energy model as the best model to fit these data, 
to the point of being considered the standard model of 
cosmology.

Other authors, obtaining equally 
good fits without considering a dark energy term (with 
more modern data samples) were more critical \cite{Nielsen2016}.
Only a year after Riess's work, there were already 
authors who performed their own study of the data by 
wondering whether the extinction due to intergalactic 
dust was sufficient without the need for dark energy \cite{Aguirre1999}.
Others considered possible time evolution of type 
Ia supernovae \cite{Banerjee2007,Dominguez1999}.
Numerous papers have also been published that examine
other cosmologies that could fit the data, including 
exotic cases such as the Carmeli cosmology \cite{Oliveira2006}, 
the Lema\^itre--Tolman--Bondi solution \cite{Romano2007}, and 
relativistic effects on the recesion of galaxies \cite{Far10}.

In this work we have focused on well-known cosmologies, as 
well as on simple models for extinction, evolution, and 
partial redshift not due to expansion.
As expected, the standard model continues to fit the data 
excellently, just as it did prior to the small SNe 
Ia samples used by Riess et al.\ and  Perlmutter et al. It is 
very interesting to see, however, that other models, discarded 
as a basis for current cosmological study, also fit the supernova data.

There are also other evidences that favor the $\Lambda $CDM model with 
$\Lambda$ non-zero over other  models \cite{Tegmark2004}; 
namely, the cosmic microwave background radiation (CMBR), 
the large-scale structure of the distribution of galaxies, 
and the measurement of the age of the oldest objects in the 
universe. Cosmologies such as the static universe and 
Einstein--de Sitter have trouble explaining some (if not all) 
of them; more theoretical developments would be needed to 
modify such cosmologies to explain these lines of evidence. It is also 
true that a standard model including dark energy has 
numerous problems and difficulties with no clear solutions in 
sight \cite{Perivolaropoulos2021,Abd22,Lop22}.
The nature of dark energy is also a mystery. Associating it 
with the energy of the quantum vacuum leads to one of the 
greatest contradictions in physics (over 120 orders of 
magnitude \cite{Carroll2001}).
In any case, the discussion of the different cosmological 
tests and the theoretical meaning of 
dark energy term are not included in the topics discussed here, but only the fit
of the Hubble-Lema\^itre diagram for SNe Ia concerning us here.

Listed below are the models discussed in this paper that fit the 
Pantheon data well ($Q$ indicates the probability), with 
values of $A$ similar to the standard model (thus, the 
value of $M$ is not discrepant with the local SNe Ia
measurements):
\begin{itemize}
  \item $\Lambda $CDM, $\Omega_M = 0.287$ ($Q = 0.678$).
  \item FLRW with curvature, $\Omega_M = 0.355$, 
$\Omega_\Lambda = 0.835$ ($Q = 0.696$).
  \item Einstein--de Sitter with extinction ($Q = 0.453$).
  \item Linear Hubble--Lema\^itre law static Euclidean with 
extinction ($Q = 0.333$).
  \item Static Euclidean with tired light with extinction ($Q = 0.275$).
  \item FLRW with curvature and with evolution 
$\Omega_M = 0.957$, $\Omega_\Lambda \approx 0$ ($Q = 0.702$).
  \item Einstein--de Sitter with evolution $\Omega_M = 1$, 
$\Omega_\Lambda = 0$ ($Q = 0.709$).
\end{itemize}

Other cosmologies with a significant non-zero probability 
(although very low and practically ruled out) and/or
 values of $A$ very different from the standard one are:
FLRW with curvature and $\Lambda =0$ ($Q=0.015$)
Milne Universe ($Q = 0.016$); and several combinations of 
cosmological model+partial non-cosmological redshift.

The most interesting result of this analysis is that we have 
that a model with the same probability as that of the standard 
model or even slightly higher is the Einstein--de Sitter 
model with linear time evolution in the absolute magnitude 
of SNe Ia, with only one free parameter other than amplitude 
$A$. Similarly, when adding evolution to $\Lambda $CDM-type 
models with free parameters, the best fit is precisely that 
with $\Omega_M \sim 1$ and $\Omega_\Lambda = 0$.
Obviously, with the values of $Q$ shown in the list, we cannot 
be sure that the Einstein--de Sitter model is confirmed by the 
data above the $\Lambda $CDM (which also has a very high $Q$ value), 
but it is clear that there is still much to be said about 
the cosmological parameter fit from distant SNe Ia data.

The inclusion of dark energy (and thus accelerated expansion of 
the universe) is not necessary in view of this analysis. 
There is degeneracy in several variables: dark energy, 
extinction, evolution, partially non-cosmological redshifts 
(although requiring calibration of $M$ far from compatibility 
with local SNe measurements), and possibly other paramaters 
that we have not explored here.

\section*{Acknowledgements}
Thanks are given to the anonymous referee for helpful comments. Thanks are given to Terry Mahoney (IAC)
for proofreading of this text.








\end{document}